\documentclass[preprint,showpacs,preprintnumbers,amsmath,amssymb]{revtex4}

\usepackage{graphicx}% Include figure files
\usepackage{dcolumn}% Align table columns on decimal point
\usepackage{amsmath}
\usepackage{amssymb}
\def\vq{{\bf q}}

\def\vk{{\bf k}}
\def\vQ{{\bf Q}}

\def\vr{{\bf r}}
\def\vS{{\bf S}}

\newcommand{\fig}[1]{Fig.~\ref{#1}}
\newcommand{\be}{\begin{equation}}
\newcommand{\ee}{\end{equation}}
\newcommand{\bea}{\begin{eqnarray}}

\newcommand{\eea}{\end{eqnarray}}
\newcommand{\bean}{\begin{eqnarray*}}
\newcommand{\eean}{\end{eqnarray*}}
\newcommand{\bfi}{\begin{figure}}
\newcommand{\efi}{\end{figure}}
\newcommand{\bc}{\begin{center}}
\newcommand{\ec}{\end{center}}
\newcommand{\ba}{\begin{array}}
\newcommand{\ea}{\end{array}}

\begin{document}

%\preprint{APS/123-QED}

\title{Theory of reduced singlet pairing without the underlying state of charge stripes in 
the high-temperature superconductor YBa$_{2}$Cu$_{3}$O$_{6.45}$}

\author{Hiroyuki Yamase} 
%\author{Hiroyuki Yamase}
% \email{h.yamase@fkf.mpg.de} 
% \altaffiliation[Also at ]{Physics Department, XYZ University.}
\affiliation{Max-Planck-Institute for Solid State Research, 
Heisenbergstrasse 1, D-70569 Stuttgart, Germany}


%\affiliation{

\date{\today}% It is always \today, today,

\begin{abstract} 
Recently, a strongly enhanced $xy$ anisotropy of magnetic excitations 
was observed in YBa$_{2}$Cu$_{3}$O$_{y}$ (YBCO$_{y}$) with 
$y=6.45$ and $T_{c}=35$ K [Science {\bf 319}, 597 (2008)].  
Unlike the observation in YBCO$_{6.6}$ and YBCO$_{6.85}$, 
the anisotropy grows to be pronounced at lower temperature 
and at lower energy, and is not suppressed by the onset of 
superconductivity. We propose that the effect of 
singlet pairing is substantially reduced in YBCO$_{6.45}$. 
This reduction 
concomitantly enhances an order competing with singlet pairing, 
a strong tendency of 
the so-called $d$-wave Pomeranchuk instability,
leading to the magnetic excitations observed experimentally. 
\end{abstract}

\maketitle

%\section{\label{sec:level1}Introduction}
%\section{Introduction} 
The spin-charge stripe order was widely discussed in high-$T_{c}$ 
cuprates.\cite{kivelson03} In this scenario, 
the two-dimensional CuO$_{2}$ plane 
is assumed to have the instability of one-dimensional 
charge order, the so-called stripe pattern, in the antiferromagnetic (AF)  
background with a $\pi$ phase shift across the charge stripe.  
However, the charge order signal was observed only for limited materials 
with specific hole doping rates and had a broad spectrum with 
very weak intensity,\cite{tranquada95,niemoller99,ichikawa00,fujita02} 
implying that the charge order is usually not well developed. 
Kivelson, Fradkin, and Emery introduced a new concept of electronic  
nematic order,\cite{kivelson98} which was envisaged as the melting of  
charge stripe order,\cite{nussinov08} 
where the translational symmetry is recovered 
but the orientational symmetry is still broken. 

On the other hand, an alternative root to yield electronic 
nematic order was found in minimal models of high-$T_{c}$ cuprates 
such as the $t$-$J$\cite{yamase00,miyanaga06,edegger06} 
and Hubbard\cite{metzner00,wegner02,honerkamp02,kampf03,neumayr03} models. 
These models show a tendency toward $d$-wave type Fermi surface 
deformations ($d$FSD), the so-called $d$-wave 
Pomeranchuk instability. 
The Fermi surface (FS) expands along the $k_{x}$ direction 
and shrinks along the $k_{y}$ direction, or vice versa. 
The $d$FSD state has the same symmetry as 
the electronic nematic state. But the physical origin is different from 
the stripe physics,\cite{kivelson98} since the $d$FSD is generated 
by forward scattering processes of quasiparticles, 
not by fluctuating charge stripes. 

The $d$FSD competes with superconductivity.\cite{yamase07a}  
In the slave-boson mean-field\cite{yamase00} and variational Monte Carlo\cite{edegger06} 
analyses of the $t$-$J$ model, superconductivity becomes dominant. 
Yet the system still has sizable $d$FSD correlations,\cite{yamase04b}  
producing a giant response to a small external anisotropy, 
e.g., due to a lattice structure 
and anisotropic strain.\cite{yamase00,edegger06} 
This theoretical insight yields a promising scenario\cite{yamase06} 
to understand the pronounced anisotropy of magnetic excitations 
observed in untwinned YBa$_{2}$Cu$_{3}$O$_{y}$ (YBCO$_{y}$) with 
$y=6.5$,\cite{stock04} $6.6$,\cite{hinkov04,hinkov07} 
and $6.85$,\cite{hinkov04} 
where the lattice yields a small $xy$ anisotropy. 

Quite recently, a very strong anisotropy was observed in magnetic 
excitations in untwinned 
YBCO$_{6.45}$.\cite{hinkov08} 
The observed spectra were qualitatively different 
from typical observations in Y-based cuprates. 
(i) The anisotropy starts to increase 
below $\sim 150$ K and saturates below $\sim 50$ K, 
in contrast to the case of 
YBCO$_{6.6}$\cite{hinkov04,hinkov07} where the anisotropy 
is most pronounced at relatively high temperature 
and is reduced at low temperature. 
(ii) The magnetic excitation spectra are hardly affected by the onset of 
superconductivity and in fact a gap feature is not found in 
the magnetic excitations. This is sharply different from the well-known 
observation in Y-based cuprates, where 
a broad spectrum around $\vq=(\pi,\pi)$ 
becomes very sharp and 
is strongly enhanced below the onset temperature 
of superconductivity, 
leading to a resonance at relatively high energy, followed by 
suppression of low energy spectral weight.\cite{fong00,dai01} 
These peculiar magnetic excitations in YBCO$_{6.45}$ 
are not captured even qualitatively by simple 
application of the theory in Ref.~\onlinecite{yamase06}, implying failure of the 
standard RVB mean-field theory in a strongly underdoped region.

Why do the qualitative features of 
magnetic excitations change so drastic in YBCO$_{6.45}$? 
Since YBCO$_{6.45}$ lies closer to the AF instability, strong 
AF fluctuations might play an important role. 
However, even for the lower hole-doped system 
YBCO$_{6.353}$, and thus much closer to the AF instability, 
the experimental data\cite{stock06} 
showed that magnetic correlation length is 
still short ranged. 
In this paper, we show an integral role of $d$FSD correlations, 
which are concomitantly enhanced by the suppression of 
singlet pairing, leading to results very similar to 
the experimental observation.\cite{hinkov08} 

%\section{Model and formalism} 
We analyze the bilayer $t$-$J$ model on a square lattice, 
a minimal model for Y-based cuprates, \cite{miscbilayer}  
\be
 H = -  \sum_{\vr,\,\vr',\, \sigma} t_{\tau} 
 \tilde{c}_{\vr\,\sigma}^{\dagger}\tilde{c}_{\vr'\,\sigma}+
   \sum_{\langle \vr,\vr' \rangle}J_{\tau} \,
 \vS_{\vr} \cdot \vS_{\vr'}  \label{tJ} 
\ee  
defined in the Fock space with no doubly occupied sites.
The operator $\tilde{c}_{\vr\,\sigma}^{\dagger}$
($\tilde{c}_{\vr\,\sigma}$) creates (annihilates) an electron with
spin $\sigma$ on site $\vr$, and $\vS_{\vr}$ is the 
spin operator. 
$J_{\tau}$ $(>0)$ is a superexchange coupling between the 
nearest neighbor sites along each direction, $\tau=x,y,z$. 
We take into account hopping amplitudes $t_{\tau}$ between 
$\vr$ and $\vr'$ up to third-nearest neighbors, and  
the direction of $\vr'-\vr$ is represented by $\tau$. 
The Hamiltonian (\ref{tJ}) is analyzed in the standard slave-boson 
method in the $t$-$J$ model by introducing the so-called 
resonating-valence-bond (RVB) mean fields. 
However, in contrast to the standard RVB theory, 
we assume the mean field of singlet pairing $\Delta$ to be zero, 
in order to mimic possibly substantial suppression of singlet pairing 
in YBCO$_{6.45}$ as we will discuss later in detail.  
The formalism and the band parameters are otherwise the same as 
those in Ref.~\onlinecite{yamase06}, where 
we successfully discussed the pronounced anisotropy of magnetic excitations 
in YBCO$_{6.85}$\cite{hinkov04} and YBCO$_{6.6}$.\cite{hinkov04,hinkov07}   
We introduce $5\%$ $xy$ anisotropy into the hopping integrals 
$t_{\tau}$ and $10\%$ into the superexchange coupling $J_{\tau}$, 
twice as large, as imposed by the superexchange mechanism. 
The anisotropy then can be parameterized by a single parameter 
$\alpha=-0.05$. 
The minus sign indicates that the band parameters are more enhanced 
along the $y$ direction, parallel to the direction of the CuO chains, 
which indeed reproduces\cite{kao05,yamase06,schnyder06} 
qualitatively the same anisotropic distribution 
of magnetic excitations as observed previously.\cite{hinkov04,hinkov07} 
Interestingly the negative sign of $\alpha$ is also implied in a spin spiral 
state to understand the anisotropy of magnetic excitations.\cite{pardini08} 

\begin{figure}
\centerline{\includegraphics[width=0.35\textwidth]{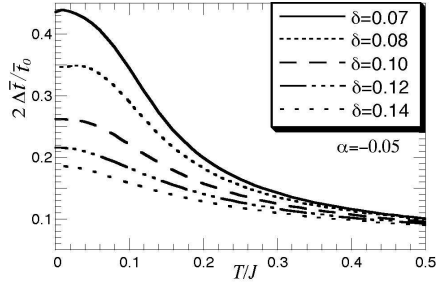}}
\caption{$T$ dependence of the anisotropy of 
renormalized band, $2\Delta \bar{t}/\Delta\bar{t}_{0}$, 
for several choices of $\delta$ for $\alpha=-0.05$.}
\label{mean-field}
\end{figure}

The most important quantity is the renormalized in-plane 
nearest-neighbor hopping 
$\bar{t}_{\tau}$, which is given by 
$\bar{t}_{\tau}=t_{\tau} \delta + \frac{3}{8} J_{\tau} \chi_{\tau}$ 
with $\tau=x, y$; $\delta$ is the hole density and $\chi_{\tau}$ is the 
RVB bond order along the $\tau$ direction. 
The order parameter of the $d$FSD is, then, defined 
as $\chi_{d}=(\chi_{x}-\chi_{y})/2$. 
Figure~\ref{mean-field} shows the temperature ($T$) dependence 
of the anisotropy of 
renormalized band $2 \Delta \bar{t}/\bar{t}_{0}$, where 
$\Delta \bar{t} = |\bar{t}_{x} - \bar{t}_{y}|/2 = |(t_{x}-t_{y}) \delta /2 
+ \frac{3}{8}(J_{x}- J_{y}) \chi_{d}|$ and 
$\bar{t}_{0} = (\bar{t}_{x} + \bar{t}_{y})/2$. 
The anisotropy is strongly enhanced at lower $T$, 
especially for lower $\delta$. 
This enhancement comes from the development of $\chi_{d}$, namely 
the underlying $d$FSD instability 
in the $t$-$J$ model.\cite{yamase00,miyanaga06,edegger06} 
The presence of the anisotropy in $t_{\tau}$ and $J_{\tau}$, however, 
smears the singularity associated with the 
$d$FSD instability, 
which would appear below 
$T \approx 0.03J (0.05J)$ for $\delta = 0.08 (0.07)$ 
and disappear for $\delta \gtrsim 0.09$ if the input anisotropy 
is zero.

The irreducible dynamical magnetic susceptibility $\chi_{0}(\vq,\omega)$ 
reads 
\be 
\chi_{0}(\vq,\,\omega) = \frac{1}{4 N} 
\sum_{\vk} \frac{\tanh \frac{\xi_{\vk}}{2T}
   -\tanh \frac{\xi_{\vk +\vq}}{2T}}
{\xi_{\vk}-\xi_{\vk+\vq}+\omega+{\rm i}\Gamma}\,,
\label{xoqw}
\ee
where $N$ is the total number of (bilayer) lattice sites and 
$\Gamma$ is a positive infinitesimal. 
We calculate magnetic excitations 
in the renormalized random phase 
approximation (RPA)\cite{brinckmann99,yamase99} 
\be
\chi(\vq,\,\omega)=\frac{\chi_{0}(\vq,\,\omega)}{1+ 
J(\vq)\chi_{0}(\vq,\,\omega)}\; . \label{RPA}
\ee
Here 
\be
J(\vq) = 2r (J_{x}\cos q_{x}+J_{y}\cos q_{y}) +J_{z}\cos q_{z} \label{RPAJ}
\ee
with a renormalization factor $r$. 
While $r=1$ in the plain RPA, the value of $r$ is reduced in 
the renormalized RPA such that the AF instability 
appears in a doping region comparable with actual materials. 
We set $r=0.4$, which confines the AF instability to 
$\delta \lesssim 0.068$. 

\begin{figure}
\centerline{\includegraphics[width=0.4\textwidth]{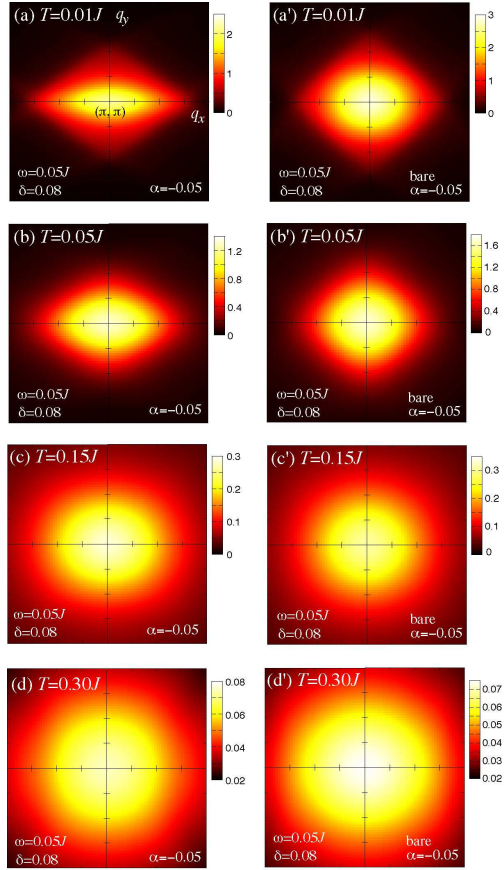}}
\caption{(Color online) Left-hand panels: 
$\vq$ maps of Im$\chi(\vq,\omega)$ for a sequence of 
$T$ in $0.6\pi \leq q_{x}, q_{y} \leq 1.4\pi$ for $\delta=0.08$, 
$\omega=0.05J$, and $\alpha=-0.05$. 
Right-hand panels: corresponding results 
for the bare anisotropy, namely without $d$FSD correlations.} 
\label{q-map}
\end{figure}

We calculate $\chi(\vq,\omega)$ numerically by choosing $\Gamma =0.05J$. 
This choice of $\Gamma$ is mainly due to numerical convenience, 
but simulates damping of electrons by static defects in real materials 
and broadening due to limited energy resolution in neutron 
scattering experiments. 
Since neutron scattering measurements for untwinned YBCO are 
performed for the odd channel, 
we focus on this channel ($q_{z}=\pi$). 
Considering that the hole density in 
YBCO$_{6.45}$ is estimated as $8.5\%$,\cite{hinkov08} we fix 
$\delta=0.08$.

%\section{Results}

The left-hand panels of \fig{q-map} show  
two-dimensional $\vq$ maps of the imaginary part 
of $\chi(\vq,\omega)$ at $\omega=0.05J$ for a sequence of temperatures. 
At $T=0.01J$, significant spectral weight is centered around 
$\vq=(\pi,\pi)$ and is strongly elongated along the $q_{x}$ direction. 
This pronounced anisotropy originates from the strong 
enhancement of the original band anisotropy due to 
$d$FSD correlations (\fig{mean-field}). 
The anisotropy of magnetic excitations 
is reduced with increasing $T$. 
In particular, the distribution of the spectral 
weight becomes nearly symmetric around $(\pi,\pi)$ at $T=0.30J$. 
To clarify the $d$FSD effect, 
we show the corresponding result 
for the {\it bare} anisotropy effect 
by switching off $d$FSD correlations in the right-hand panels of 
\fig{q-map}, where we impose the same 
anisotropy $\alpha=-0.05$, but assume $\chi_{d} \equiv 0$. 
Although the value of $\Delta\bar{t}$ does not depend on $T$, we see that 
the anisotropy of Im$\chi(\vq,\omega)$ depends weakly on $T$ and 
becomes visible with decreasing $T$. 
But the obtained anisotropy [Figs.~\ref{q-map}(a') and \ref{q-map}(b')] 
is much weaker than the results due to $d$FSD 
correlations  [Figs.~\ref{q-map}(a) and \ref{q-map}(b)].  
Hence the underlying $d$FSD correlations are particularly important to 
produce a strongly enhanced anisotropy at lower $T$.

\begin{figure}
\centerline{\includegraphics[width=0.35\textwidth]{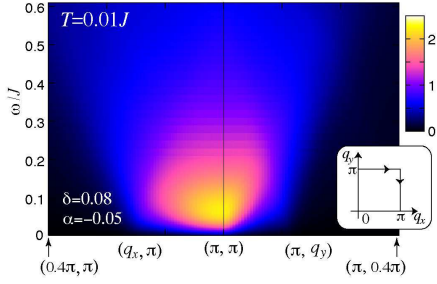}}
\caption{(Color online) 
$(\vq,\omega)$ map of Im$\chi(\vq,\omega)$ at $T=0.01J$ 
for $\delta=0.08$ and $\alpha=-0.05$; 
the $\vq$ directions are shown in the inset.  
}
\label{qw-map}
\end{figure}

To see the $\omega$ dependence of the anisotropy, 
the $(\vq, \omega)$ map of Im$\chi(\vq,\omega)$ 
is shown in \fig{qw-map} at low $T$.   
While the anisotropic distribution 
is still discernible for a high energy where 
the spectral intensity is substantially reduced, 
a pronounced anisotropy appears especially in a low energy region.

\begin{figure}
\centerline{\includegraphics[width=0.35\textwidth]{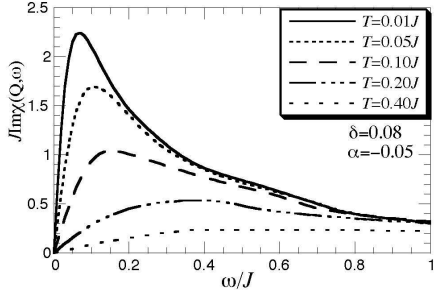}}
\caption{
$\omega$ dependence of Im$\chi(\vQ,\omega)$ for several choices of 
$T$ for $\delta=0.08$ and $\alpha=-0.05$.  
}
\label{w-scan}
\end{figure}

The evolution of the spectral intensity of Im$\chi(\vq,\omega)$ 
is shown in \fig{w-scan} 
as a function of $\omega$ for several choices of $T$ 
at $\vq=\vQ=(\pi,\pi)$. 
At high $T$, Im$\chi(\vQ,\omega)$ shows 
a broad maximum at a moderate energy. 
With decreasing $T$, low-energy spectral weight is 
substantially enhanced, yielding a peak structure at low energy.

%\section{Discussion}
When $d$FSD correlations strongly enhance a small $xy$ anisotropy 
of the original 
band dispersion, a strong anisotropy of Im$\chi(\vq,\omega)$ 
is induced with a characteristic $T$ dependence. 
In YBCO$_{6.45}$, the experimental data\cite{hinkov08} suggest that 
fermions responsible for magnetic excitations feel a rather small 
mean field of singlet pairing in the sense that neither a gap 
feature nor a clear impact of the onset of superconductivity is observed 
in neutron scattering. 
This special experimental situation is 
mimicked by setting $\Delta \equiv 0$ in the slave-boson 
scheme of the $t$-$J$ model. We have found that 
this phenomenological treatment well 
captures the most salient features 
observed in YBCO$_{6.45}$:\cite{hinkov08}  
(i) the strongly enhanced anisotropy of Im$\chi(\vq,\omega)$ at lower 
$T$ for low $\omega$ (Figs.~\ref{q-map} and ~\ref{qw-map})  
and (ii) the enhanced spectral weight of 
Im$\chi(\vQ,\omega)$ at low $\omega$ for low $T$ (\fig{w-scan}).

On the other hand, 
in the standard slave-boson mena-field theory of 
the $t$-$J$ model\cite{yamase06}, the onset temperature of singlet 
pairing is higher than that of the $d$FSD, and 
the singlet formation suppresses 
the susceptibility of the $d$FSD, 
which then does not diverge.\cite{yamase00} 
The susceptibility, however, remains 
appreciable\cite{yamase04b} so that 
a small external anisotropy introduced in the $t$-$J$ model 
is strongly enhanced 
even in the singlet 
pairing state.\cite{yamase00} 
This effect is more pronounced for a lower doping rate. 
Such an enhanced anisotropy nonetheless turns out not to be sufficient to 
capture the anisotropy of Im$\chi(\vq,\omega)$ observed 
in YBCO$_{6.45}$\cite{hinkov08} even at qualitative level for 
a realistic choice of parameters. In particular, 
the anisotropy is enhanced at relatively high $T$  
(Fig.~20 in Ref.~\onlinecite{yamase06}). 
Moreover the experimental observation\cite{hinkov08} that the onset of 
superconductivity hardly affects magnetic excitations is 
difficult to be captured in existing microscopic theories for cuprates. 

Inclusion of singlet pairing is not so straightforward, 
and something beyond the existing theories happens in YBCO$_{6.45}$. 
Here focusing on the context of the present paper we discuss several 
possibilities that $d$FSD correlations still 
remain strong enough to be compatible with the present result 
even in a full calculation. 
First, since the $d$FSD instability concerns discrete symmetry breaking, 
the spontaneous symmetry breaking is allowed at finite temperature 
even in the exact   
analysis of the two-dimensional model, 
in contrast to superconductivity and antiferromagnetism. 
Hence suppression of the $d$FSD due to fluctuations is less crucial 
than the latter two, in favor of strong correlations of the $d$FSD. 
In the presence of a small external $xy$ anisotropy, then, 
the anisotropy can be strongly enhanced as we have 
seen in \fig{mean-field}. 
The feedback of the enhanced anisotropy appears as further reduction 
of the magnitude of singlet pairing compared with the mean-field 
value in the slave-boson theory of the $t$-$J$ model.  
Second, YBCO$_{6.45}$ has an orthorhombic crystal structure, but is 
close to the continuous transition to a tetragonal structure. 
The $d$FSD order parameter couples to a phonon related to the 
orthorhombic-tetragonal structural transition, which 
contributes to an enhancement of $d$FSD correlations. 
In this case, the phonon spectrum may also show  
strong anisotropy in the orthorhombic phase. 
Last, there might be a phase segregation into hole-poor 
AF domains with nano-scale correlation lengths   
and hole-rich superconducting islands in 
YBCO$_{6.45}$, and the neutron scattering signals 
originate mainly from the AF domains. 
Such finite-sized domains are often discussed in a strongly 
underdoped superconducting region 
in $\mu$SR measurements.\cite{niedermayer98,sanna04}

While the present result well captures the neutron scattering data 
in YBCO$_{6.45}$,\cite{hinkov08} a detailed comparison 
reveals disagreement about several aspects, which may be mainly due to 
our simple calculation ignoring singlet pairing. 
(i) A weakly incommensurate 
structure was observed along the $q_{x}$ direction, whereas 
we have obtained a single peak at $(\pi,\pi)$ as read off from 
\fig{q-map}. In a fermiology scenario, the incommensurate structure in 
Y-based cuprates is explained by the suppression of the commensurate 
peak due to 
the development of $d$-wave singlet pairing.\cite{zha93,yamase06} 
Hence the observed weak incommensurate peak can be 
due to the development of relatively weak singlet pairing. 
(ii) The observed data at $\omega = 50$ meV was interpreted to be 
isotropic.\cite{hinkov08} In the present calculation, the anisotropy 
becomes less pronounced for high $\omega$, but is still 
discernible (\fig{qw-map}). 
On the other hand, if we include singlet pairing in the slave-boson 
mean-field theory, it is known\cite{yamase06} that 
the anisotropy of Im$\chi(\vq,\omega)$ at high $\omega$ 
depends strongly on energy. 
More detailed experiments as well as more complete calculations are 
necessary to discuss the anisotropy at high $\omega$.

%\section{Conclusion}
The present phenomenological 
calculation suggests that $d$-wave 
singlet pairing is substantially reduced in YBCO$_{6.45}$, which 
then concomitantly enhances strong $d$FSD correlations, 
leading to a magnetic excitation spectrum  
very similar to the experimental data.\cite{hinkov08} 
Since the magnitude of $d$-wave singlet pairing is often discussed 
to become larger with decreasing doping rate, 
the present study implies an interesting direction 
to understand the evolution of $d$-wave singlet pairing  
as well as its connection with $d$FSD correlations 
in underdoped cuprates. 
Recent angle-resolved photoemission spectroscopy data\cite{tanaka06} 
is in fact suggestive of relatively small $d$-wave 
singlet pairing. 
It demonstrated two distinct gaps, 
the amplitude of one of which increases for lower doping rates 
while the other, which shows $d$-wave symmetry, decreases. 
In addition, strong $d$FSD fluctuations substantially reduce 
the lifetime of quasiparticles 
in the antinodal region of the FS while not in the nodal 
direction as shown in Ref.~\onlinecite{metzner03}. 
Hence strong $d$FSD fluctuations 
may play an integral role to understand the pseudogap 
in the strongly underdoped regime of YBCO. 
Strong $d$FSD fluctuations are also expected in a 
scenario of a quantum phase transition into the $d$FSD state 
in a underdoped region.\cite{sachdev08}  
Such a scenario is, however,  based on the assumption of the existence of robust  
$d$-wave pairing, which is different from the present theory where 
we have invoked  
the substantial suppression of $d$-wave pairing due to the competition\cite{yamase04b,yamase07a} 
with the $d$FSD in a strongly underdoped region. 

%\begin{acknowledgments}
The author is grateful to V. Hinkov and B. Keimer for sharing 
their unpublished data with him and to W. Metzner and A. Toschi 
for very helpful discussions. 
%\end{acknowledgments}

%\appendix*

%\newpage
\bibliography{main.bib}% Produces the bibliography via BibTeX.

\begin{thebibliography}{37}
\bibitem[{mis()}]{miscbilayer}
\bibinfo{note}{Since the $q_z$ modulation of magnetic excitations observed in
  YBCO is known to be well characterized by a bilayer model, we do not consider
  a direct coupling to CuO chains.}
\end{thebibliography}

\end{document}